\documentstyle[sprocl]{article}
\newcommand{\beq}{\begin{equation}}
\newcommand{\eeq}{\end{equation}}
\newcommand{\beqa}{\begin{eqnarray}}
\newcommand{\eeqa}{\end{eqnarray}}
\newcommand{\ba}{\begin{array}}
\newcommand{\ea}{\end{array}}

\begin{document}

\title{NEW RESULTS ON QUANTUM CHAOS IN ATOMIC NUCLEI}

\author{V.R. MANFREDI}

\address{Dipartimento di Fisica "G. Galilei"\\
Universit\`a di Padova, Via Marzolo 8, I--35131\\ 
Istituto Nazionale di Fisica Nucleare \\
Sezione di Padova, Via Marzolo 8, I--35131 Padova, Italy}

\author{L. SALASNICH}

\address{Dipartimento di Matematica Pura ed Applicata \\
Universit\`a di Padova, Via Belzoni 7, I--35131 Padova, Italy\\
Istituto Nazionale di Fisica Nucleare \\
Sezione di Padova, Via Marzolo 8, I--35131 Padova, Italy}

\maketitle\abstracts{In atomic nuclei, 
ordered and chaotic states generally coexist. 
In this paper the transition from ordered to chaotic states will be discussed 
in the framework of roto--vibrational and shell models. 
In particular for $^{160}Gd$, in the roto--vibrational model, the Poincar\`e 
sections clearly show the transition from order to chaos for different 
values of rotational frequency. Furthermore, the spectral statistics of 
low--lying states of several $fp$ shell nuclei are studied with realistic 
shell--model calculations.}

\section{Introduction}

In atomic nuclei, just as in other many--body systems, 
ordered and chaotic states
\footnote{Here we use the term chaos for a quantal system in a very 
restricted sense: "{\it Quantum Chaology is the study of semiclassical 
behaviour characteristics of systems whose classical motion exhibits chaos}". 
If the semiclassical limit is not known we use spectral statistics$^{4)}$ 
(e.g. $P(s)$ and $\Delta_3(L)$) to distinguish between 
ordered and chaotic states.} 
generally coexist. 
In fact, in zero--order approximation the relevant 
elementary excitations (such as rotation, vibration and single--particle) 
may be regarded as independent modes. Then we consider the interaction between 
these elementary modes$^{1)}$. At one end of the chain of complexity 
we have a single mode which can be considered "regular", 
whereas at the other end there are the so--called "stochastic" 
or "chaotic states". 
\par
In order to discuss the coexistence in atomic nuclei of ordered and chaotic 
states, many models have been used$^{2),3)}$ but, for reasons of space, we 
shall mention only the roto--vibrational and shell models. 

\section{The Roto--Vibrational Model}

This model has been amply described in Ref. 5, 6 and 7 and we 
limit ourselves to reporting only a few basic formulae. 
The hamiltonian is
\beq
H=H_{vib}+H_{rot},
\eeq
where
\beq
H_{vib}={1\over 2}B({\dot a}_0^2+2{\dot a}_2^2)+V(a_0,a_2),
\eeq
\beq
H_{rot}={1\over 2}\sum_{k=1}^3 \omega_k^2 J_k(a_0,a_2),
\eeq
with
\beq
V(a_0,a_2)= {1\over 2}C_2(a_0^2+2a_2^2)+
\sqrt{2\over 35}C_3a_0(6a_2^2-a_0^2)+{1\over 5}C_4(a_0^2+2a_2^2)^2+V_0.
\eeq
The shape of the nuclear potential $V(a_0,a_2)$ is function of $C_2$ 
and $\chi =C_3^2/(C_2 C_4)$$^{7)}$. 
The parameters $a_0$ and $a_2$ are connected to the deformation 
$\beta$ and asymmetry $\gamma$ by the relations
\beq
a_0=\beta \cos{\gamma}, \;\;\; a_2={\beta\over \sqrt{2}}\sin{\gamma}.
\eeq
In terms of the new variables the components of the moment of inertia 
are$^{8)}$ 
\beq
J_k=4B\beta^2 \sin^2{(\gamma -{2\pi\over 3}k)}.
\eeq 
If the nucleus has an axially symmetric equilibrium configuration and
\beq
\omega_1=\omega_2={\omega\over \sqrt{2}}, \;\;\; \omega_3=0,
\eeq
the hamiltonian (1) can be written as
\beq
H={1\over 2}B({\dot a}_0^2+2{\dot a}_2^2)+V(a_0,a_2)
+{1\over 2}B\omega^2 (3a_0^2+2a_2^2),
\eeq
where $V(a_0,a_2)$ is given by (4) and $V_0$ is chosen to have 
the minimum of the potential equal to zero. 

\section{Numerical study of the order--chaos transition}

A very useful tool for the study of the global instability is provided by the 
Poincar\`e sections. The classical trajectories have 
been calculated by a fourth order Runge--Kutta method$^{9)}$. 

\begin{figure}
\vskip 10. truecm
\caption{The Poincar\`e sections for $^{160}Gd$ at the energy 
$5.5$ MeV and for different values of rotational frequency; 
from the top: $\hbar \omega=0$ MeV, $\hbar \omega=0.5$ MeV, 
$\hbar \omega =1$ MeV. ~~~~~~~~~~~~~~~~~~~~~~~~~~~~~~~~~~~~~~
Adapted from Ref. 9.}
\vskip 0.5 truecm
\end{figure} 

\par
The Hamilton equations are as follows
$$
{\dot a_0}=B p_{0}, 
$$
$$
{\dot a_2}=2B p_{2},
$$
\beq
{\dot p_{0}}=-C_2a_0-2\sqrt{2\over 35}C_3(3a_2^2-3a_0^2)-
{4\over 5}C_4a_0(a_0^2+2a_2^2)-3B\omega^2a_0,
\eeq
$$
{\dot p_{2}}=-2C_2a_2-12\sqrt{2\over 35}C_3a_0a_2-
{8\over 5}C_4a_2(a_0^2+2a_2^2)-2B\omega^2a_2,
$$
where $p_0$ and $p_2$ are the conjugate momenta 
\beq
p_{0}=B\dot{a_0}, \;\;\;\; p_{2}=2B\dot{a_2}.
\eeq

\begin{figure}
\vskip 10. truecm
\caption{The Poincar\`e sections for $^{166}Er$ at the 
rotational frequency $\omega =0$ and for different values of the energy: 
(a) $E=1$ MeV, (b) $E=6$ MeV, (c) $E=9$ MeV, (d) $E=12$ MeV. 
~~~~~~~~~~~~~~~~~~~~~~~~~~~~~~~~~~~~~~~Adapted from Ref. 9.}
\vskip 0.5 truecm
\end{figure} 

Figure 1 shows the Poincar\`e sections for $^{160}Gd$ at the energy 
$5.5$ MeV and for different values of rotational frequency. The figure 
clearly shows a chaos--order transition as the frequency $\omega$ 
increases. In Figure 2, for $^{166}Er$, the Poincar\`e sections are shown 
for different values of the energy and rotational frequency $\omega =0$. 
As can be seen, there is a chaos--order transition, but not so sharp 
as in the previous case.
\par
Fluctuation properties of 
quantal systems with underlying classical chaotic behaviour and 
time--reversal symmetry are in agreement with the predictions of 
the Gaussian Orthogonal Ensemble (GOE), and quantum analogs of 
classically integrable systems 
display the characteristics of Poisson statistics$^{10)}$.
In general, various statistics may be used to show the local correlations 
of the energy levels; we shall discuss $P(s)$ and $\Delta_3(L)$ only. 
$P(s)$ measures the probability that two neighbouring eigenvalues are a 
distance "s" apart. For GOE we have the Wigner distribution
\beq
P(s) = {\pi \over 2} s \exp{[-{\pi\over 4}s^2]},
\eeq
which gives level repulsion. $\Delta_3(L)$ is defined for a fixed interval 
$(-L/2,L/2)$ as the least--square deviation of the staircase function $N(E)$ 
from the best straight line fitting it
\beq
\Delta_{3}(L)={1\over L}\min_{A,B}\int_{-L/2}^{L/2}[N(E)-AE-B]^2 dE,
\eeq
where $N(E)$ is the number of levels between E and zero for positive
energy, between $-E$ and zero for negative energy. $\Delta_{3}(L)$ 
provides a measure of the degree of rigidity of the 
spectrum: for a given interval L, the smaller $\Delta_{3}(L)$ is,
the stronger is the rigidity, signifying the long--range
correlations between levels. For this statistics in the GOE ensemble
\beq
\Delta_{3}(L)=\cases{{L\over 15}, \quad L \ll 1 \cr 
              \noalign{\vskip 16 truept}
            {1\over \pi^2}\ln{L},\quad L \gg 1 \cr }.
\eeq

\begin{figure}
\vskip 8. truecm
\caption{Spectral statistics $P(s)$ and $\Delta_3(L)$ for $^{166}Er$ 
at the rotational frequency $\omega =0$ for different energy regions: 
$2\leq E\leq 6$ MeV (below) and for $13\leq E\leq 17$ MeV (above).  
The solid line is the GOE statistic curve and the dashed line is the 
Poisson one. ~~~~~~~~~~~~~~~~~~~~~~~~~~~~~~~~~~~~~~~~~~~~Adapted from Ref. 9.}
\vskip 0.5 truecm
\end{figure} 

In Figure 3 the spectral statistics 
$P(s)$ and $\Delta_{3}$ are plotted for $^{166}Er$. These 
statistics confirm the classical results: for energies above 
the saddle energy ($\sim 4$ MeV) there is 
prevalently chaotic behaviour; for higher energies 
there is mixed behaviour. 

\section{Shell Model Calculations}

In this section we discuss the statistical analysis of the shell--model 
energy levels in the $A=46$--$50$ region. Exact calculations 
are performed in the ($f_{7/2}$,$p_{3/2}$,$f_{5/2}$,$p_{1/2}$) 
shell--model space, assuming $^{40}$Ca as an inert core$^{11)}$. 
Diagonalizations are performed in the {\it m}--scheme using a fast 
implementation of the Lanczos algorithm with the code ANTOINE. 
For a fixed number of valence protons and neutrons 
we calculate the energy spectrum for 
projected total angular momentum $J$ and total isospin $T$. 
The interaction we use is a minimally modified Kuo--Brown 
realistic force with monopole improvements. 
\par
We calculate the $T=T_z$ states from $J=0$ to $J=9$ 
for all the combinations of $6$ active nucleons, 
i.e. $^{46}$V, $^{46}$Ti, $^{46}$Sc, and $^{46}$Ca, and also 
for $^{48}$Ca and $^{50}$Ca. 
\par
Since we are looking for deviations from chaotic features, we are 
mainly interested in the low--lying levels, up to a few MeV above the $JT$ 
yrast line. Let us consider the energy levels up to $4$, $5$ and $6$ MeV above 
the yrast line, and calculate the fluctuations around the 
average spacing between neighboring levels. In this range of energies, 
the level spectrum can be mapped into unfolded levels with 
quasi--uniform level density by using the constant temperature 
formula. To guarantee that the results up to different energies 
are unaffected by the unfolding procedure, the unfolding is 
performed using always the whole set of levels up to 6 MeV, 
for each $JT$ set in the nucleus. 
\par
The mean level density can be assumed to be of the form 
\beq
{\bar \rho}(E)={1\over T}\exp{[(E-E_0)/T]} ,
\eeq
where $T$ and $E_0$ are constants. For fitting purposes it is better 
to use not ${\bar \rho}(E)$ but its integral ${\bar N}(E)$. We write 
\beq
{\bar N}(E)=\int_0^E {\bar \rho}(E')dE' + N_0 = \exp{[(E-E_0)/T]}-
\exp{[-E_0/T]}+N_0 .
\eeq
The constant $N_0$ represents the number of levels with energies less 
than zero. We consider this function as an 
empirical function to fit the data and let $N_0$ take non--zero values. 
The parameters $T$, $E_0$ and $N_0$ 
that best fit $N(E)$ are obtained by minimizing the function 
\beq
G(T,E_0,N_0)=\int_{E_{min}}^{E_{max}} [N(E)-{\bar N}(E)]^2 dE ,
\eeq
where $N(E)$ is the number of levels with energies less than or equal to $E$. 
The energies $E_{min}$ and $E_{max}$ are taken as the first and last 
energies of the level sequence. 
\par
As previously discussed, the spectral statistic $P(s)$ may be used 
to study the local fluctuations of the energy levels. 
$P(s)$ is the distribution of nearest--neighbour spacings 
$s_i={\tilde E}_{i+1}-{\tilde E}_i$ of the unfolded levels. 
\par
For quantum systems whose classical analogs are integrable, 
$P(s)$ is expected to follow the Poisson limit, i.e. 
$P(s)=\exp{(-s)}$. On the other hand, 
quantal analogs of chaotic systems exhibit the spectral properties of 
GOE with $P(s)= (\pi / 2) s \exp{(-{\pi \over 4}s^2)}$. 

\begin{figure}
\vskip 8. truecm
\caption{$P(s)$ distribution for low--lying levels of $fp$ shell 
nuclei with $0\leq J\leq 9$: 
(a) $^{46}V$, $^{46}Ti$ and $^{46}Sc$; 
(b) $^{46}Ca$, $^{48}Ca$ and $^{50}Ca$. The dotted, dashed and solid curves 
stand for GOE, Poisson and Brody distributions, respectively. 
Adapted from Ref. 11.}
\vskip 0.5 truecm
\end{figure}

\par
To quantify the chaoticity of $P(s)$ in terms of a parameter, 
it can be compared to the Brody distribution, 
\beq
P(s,\omega)=\alpha (\omega +1) s^{\omega} \exp{(-\alpha s^{\omega+1})},
\eeq
with 
\beq
\alpha = (\Gamma [{\omega +2\over \omega+1}])^{\omega +1}.
\eeq
This distribution interpolates between the Poisson distribution ($\omega =0$) 
of integrable systems and the GOE distribution ($\omega =1$). 
The parameter $\omega$ can be used as a simple 
quantitative measure of the degree of chaoticity. 
\par
The number of $J=0$--$9$ spacings below $4$, $5$ and $6$ MeV range 
from $42$, $66$ and $105$ in $^{46}$Ca, to $86$, $149$ and $231$ in 
$^{46}$Ti, respectively. 
\par
To obtain a better estimate of the Brody parameter, we can combine 
spacings of different nuclei. 
The number of level spacings is now sufficiently large to yield 
meaningful statistics and we see that Ca isotopes are not very 
chaotic at low energy, in contrast to other nuclei in the same region 
(Figure 4). 
\par 
Why are Ca isotopes less chaotic than their neighbors? 
We observe that the two--body matrix elements of the 
proton--neutron interaction are, on average, larger than those of 
the proton--proton and neutron--neutron interactions. 
Consequently the single--particle mean--field motion in nuclei with both 
protons and neutrons in the valence orbits suffers 
more disturbance and is thus more chaotic. 

\section{Conclusions}

We have shown that in the roto--vibrational model of atomic nuclei, 
an order--chaos--order transition occurs as a function of the energy. 
Concerning the shell model calculations, the main conclusion of this 
paper is that for $Ca$ isotopes we find significant deviations from the 
predictions of the random--matrix theory. 

\section*{References}

\begin{description}

\item{\ 1.} B.R. Mottelson, Elementary Excitation in the Nucleus, {\it 
Proc. Int. Sum. School E. Fermi} LXIX (1977).

\item{\ 2.} O. Bohigas and H.A. Weidenm\"uller, 
{\it Ann. Rev. Nucl. Part. Sci.} {\bf 38}, 421 (1988).

\item{\ 3.} M.T. Lopez--Arias, V.R. Manfredi and L. Salasnich, 
{\it La Rivista del Nuovo Cimento} {\bf 17}, n. 5 (1994).

\item{\ 4.} M.V. Berry, Dynamical Chaos, 
{\it Proc. Roy. Soc. Dis. Meeting} (1987).

\item{\ 5.} J.M. Eisenberg and W. Greiner,  
{\it Nuclear Models}, vol. 1 (North Holland, Amsterdam, 1970).

\item{\ 6.} A. Faessler and  W. Greiner, 
{\it Zeit. Phys.} {\bf 168}, 425 (1962); 
A. Faessler, W. Greiner and R.K. Sheline, {\it Nucl. Phys.} {\bf 70}, 
33 (1965).

\item{\ 7.} U. Mosel and W. Greiner, {\it Zeit. Phys.} {\bf 217}, 256 (1968).

\item{\ 8.} A. Bohr and B. Mottelson, 
{\it Nuclear Structure}, vol. 2 (Benjamin, London, 1975).

\item{\ 9.} V.R. Manfredi and L. Salasnich,  
{\it Int. J. Mod. Phys.} E {\bf 4}, 625 (1995). 

\item{\ 10.} V.R. Manfredi, M. Rosa--Clot, L. Salasnich and S. Taddei,  
{\it Int. J. Mod. Phys.} E {\bf 5}, N. 3 (1996).

\item{\ 11.} E. Caurier, J.M.G. Gomez, V.R. Manfredi and L. Salasnich,  
{\it Phys. Lett.} B {\bf 365}, 7 (1996).

\end{description}

\end{document}